\documentclass[preprint,12pt, a4paper]{elsarticle}

\usepackage{amssymb}
\usepackage{amsthm}
\usepackage{amsmath}
\usepackage[colorlinks=true,linkcolor=blue,citecolor=blue]{hyperref}
\usepackage[final]{changes}
\usepackage[flushleft]{threeparttable}
\usepackage{colortbl}

\usepackage{float}
\restylefloat{table}

\journal{SoftwareX}

\begin{document}

\begin{frontmatter}



\title{A Python-based tool for constructing observables from the DSN's closed-loop archival tracking data files}


\author[1,2]{Ashok Kumar Verma}

\address[1]{Department of Earth, Planetary, and Space Sciences, University of California, Los Angeles, CA 90095, USA}
\address[2]{NASA Goddard Space Flight Center, Greenbelt, MD-20771, USA}

\begin{abstract}

Radio science data collected from NASA's Deep Space Networks (DSNs) are made available in various formats through NASA's Planetary Data System (PDS). The majority of these data are packed in complex formats, making them inaccessible to users without specialized knowledge. In this paper, we present a Python-based tool that can preprocess the closed-loop archival tracking data files (ATDFs), produce Doppler and range observables, and write them in an ASCII table along with ancillary information. ATDFs are primitive closed-loop radio science products with limited available documentation. Early in the 2000s, DSN deprecated ATDF and replaced it with the Tracking and Navigation Service Data Files (TNF) to keep up with the evolution of the radio science system. Most data processing software (e.g., orbit determination software) cannot use them directly, thus limiting the utilization of these data. As such, the vast majority of historical closed-loop radio science data have not yet been processed with modern software and with our improved understanding of the solar system. The preprocessing tool presented in this paper makes it possible to revisit such historical data using modern techniques and software to conduct crucial radio science experiments. 

\end{abstract}

\begin{keyword}
radio science \sep ATDF \sep closed-loop \sep DSN



\end{keyword}

\end{frontmatter}

\section*{Code metadata}
\label{metadata}

\begin{table}[H]
\begin{tabular}{lp{6.5cm}p{6.5cm}}
\hline
& Current code version & v01 \\

&Permanent link to code/repository used for this code version & \url{https://github.com/ashokverma24/atdf2ascii.git} \\

&Legal Code License   & MIT License \\

&Code versioning system used & Git \\

&Software code languages, tools, and services used & Python \\

&Compilation requirements, operating environments & bitstring, binascii, termcolor \\

&Support email for questions & ashokkumar.verma@nasa.gov\\
\hline
\end{tabular}
\label{} 
\end{table}


\section{Introduction}
\label{}

The radio signals between spacecraft and Earth-based stations enable us to conduct crucial science experiments. The technique of precisely measuring the properties of radio signals provides a distinct advantage for obtaining information about solar system bodies, including the atmosphere\cite{Fjeldbo71, Kliore02, Verma20}, gravity field\cite{kono99, Verma16, Iess18}, rotational state\cite{Anderson01, kono14, Verma16}, interior structure\cite{Folkner18, Verma18, Asmar21}, and ephemerides \cite{VermaThesis}. Furthermore, radio science observations are crucial for studying solar dynamics, including solar wind and the solar corona \cite{Verma13}, as well as testing fundamental physics\cite{Bertotti03, Verma14}. NASA's Planetary Data System (PDS) actively archives these observations and makes them publicly available to scientists for analysis. However, most of these Deep Space Network (DSN) data are archived in very complex formats that bound the usage of data outside the radio science experts.

The paper focuses on the archival tracking data files (ATDFs), which are the earliest closed-loop radio science products packed in the DSN's TRK-2-25 format\cite{atdf}. The ATDF format was used to archive crucial primitive radio science data from several NASA missions (e.g., Magellan, Galellio). Given the current advancement in computing power, a better understanding of our solar system, and continuous evolution in precise orbit determination software, making it worthwhile to revisit such historical data with modern tools (e.g, MONTE \cite{Monte}, GINS \cite{Gins}).  However, ATDF data cannot be directly accessed by most data processing software due to its complex format. As such, preprocessing software is required to unpack the binary fields of the ATDF data and convert them into useful observables. To our knowledge, there are no open-source tools available for preprocessing such closed-loop radio science data, which limits their useability.

Intending to address this need, we developed a Python-based tool to preprocess ATDF data and convert them into Doppler and range observables in accordance with the Moyer formalism\cite{Moyer}. The tool was written in Python and tested on several operating systems to maximize its usability.

\section{Formulation and software description}
\label{}

We developed software using the widely adopted Python programming language. All ATDF formatted files archived since 1986 can be preprocessed with this software. It unpacks the densely packed binaries in accordance with the TRK-2-25 interface specification\cite{atdf} and preprocesses them using Moyer's formalism\cite[Chapter 13]{Moyer}. The software currently supports high/low rate Doppler, range, and programmable frequency (ramp) data types. 

\subsection{Doppler data}
\label{Doppler}

Among the types of data available in an ATDF format, Doppler is the most common. It contains a time history of high- and/or low-rate Doppler counts, typically at the rate of 1-10 times per second. In accordance with user-provided Doppler count time (see Section \ref{functions}), the software computes the high-precision Doppler counts and converts them to Doppler shift observables. A number of radio science investigations can be conducted by using these observables, such as determining gravity fields, probing interior structure, rotational dynamics, and measuring ephemerides of spacecraft. The software creates three-way, two-way, and one-way Doppler observables as follows:

\begin{equation}
\label{dopplerObs}
f_{observable}(t) = \frac{f_{cb}}{|f_{cb}|} \times \bigg[\frac{D_{count}(t_2)-D_{count}(t_1)}{T_c} - f_{bias}(t)\bigg] \ \ \ Hz
\end{equation}

where, $f_{observable}$ is the Doppler observable at time $t$, $f_{cb}$ is the constant bias frequency, $T_c$ is the user provided count time, $D_{count}$ is the Doppler count, 
$t_2$ = t + 0.5*T$_c$, $t_1$ = t - 0.5*T$_c$, and $f_{bias}$ is the bias frequency at time $t$. 
The bias frequency, $f_{bias}$, is calculated as follows \cite{Moyer}:

\begin{equation*}
\label{bias}
f_{bias}(t) =  M2 * f_{ref}(t) - C2 * f_{transponder} + f_{cb}  \textit{  Hz,} \ \ \mbox{ for one-way Doppler}
\end{equation*}
\begin{equation}
\label{bias}
f_{bias}(t) =  f_{cb}  \textit{  Hz,} \ \ \mbox{ for two-way and \added{ three-way} Doppler}
\end{equation}

where, $f_{ref}$ is the \added{{sky-level}} reference Doppler frequency at time $t$, $M2$ is the spacecraft transponder ratio \citep{dsn201}, and $C2$ is the spacecraft downlink frequency. The reference Doppler frequency, $f_{ref}$, is calculated as follows \added{{\cite[Chapter 13, section 13.2]{Moyer}}}:

For S-band:
\begin{equation}
\label{refS}
f_{ref}(t) =  96 \times f_{osc}(t)  \textit{  Hz,} 
\end{equation}

For X-band:
\begin{equation}
\label{refX}
f_{ref}(t) =  32 \times f_{osc}(t)  + 6.5 \times 10^{9}\textit{  Hz,} 
\end{equation}

For X-band, 34-meter high efficiency DSN antennas (15, 45, and 65):
\begin{equation}
\label{refX2}
f_{ref}(t) =  32 \times (4.68125 \times f_{osc}(t) - 81.4125 \times 10^{6})   + 6.5 \times 10^{9}\textit{  Hz,} 
\end{equation}

where, $f_{osc}$ is the reference oscillator frequency. 

\added{{For Ka-band:}}
\begin{equation}
\label{refKa}
\added{{f_{ref}(t) =  1000 \times f_{osc}(t)  + 1.0 \times 10^{10} \textit{  Hz,} }}
\end{equation}

\subsection{Range data}
\label{range}
The range is another data type that is occasionally available in the ATDF. It gives direct access to the spacecraft's position relative to the ground and provides the most powerful constraints for constructing planetary ephemerides\cite{VermaThesis}. Range observables can be one-way or two-way, and they are created as follows:

\begin{equation}
\label{rangeObs}
R_{observable} =  R_{msr} - RE_{delay} + Z_{corr} - SC_{delay}, \ \ \textit{ RU}
\end{equation}

where, $R_{msr}$ is the range measurement, $RE_{delay}$ is the ranging equipment delay, $Z_{corr}$ is the \added{{DSN antenna}} Z-correction, and $SC_{delay}$ is the spacecraft delay. Range observables are usually given in Range Units (RU). The conversions between RU to seconds are given in \added{{\cite[Chapter 13, section 13.5.2]{Moyer}. Although the conversion of RU into meters depends on the signal frequency, the RU can roughly be approximated as 1 meter $\approx$ 3 RU.}}

\subsection{Ramp data}
\label{ramp}

\added{{The ATDF file also provides the time history of DSN transmitted frequencies. The transmitted frequency can be constant or time-varying (ramped).}}
When the receiver is ramped, the software also creates a ramp history file. It contains information about the ramp start and end times, the reference frequency at the beginning of the ramp, the rate at which the ramp frequency changes, the uplink band, and the corresponding DSN.

When the ramp start frequency is not known at the sky level, it is computed by Equations \ref{refS} and \ref{refX} for the S-band and X-band, respectively, and the ramp rate is determined by the first derivative of these equations:

For S-band:
\begin{equation}
\label{rampS}
f_{rate}(t) =  96 \times \dot{f}_{osc}(t)  \textit{  Hz/sec,} 
\end{equation}

For X-band:
\begin{equation}
\label{rampX}
f_{rate}(t) =  32 \times \dot{f}_{osc}(t)  \textit{  Hz/sec,} 
\end{equation}

\added{{For Ka-band:}}
\begin{equation}
\label{rampKa}
\added{{f_{rate}(t) =  1000 \times \dot{f}_{osc}(t)  \textit{  Hz/sec,} }}
\end{equation}

where, $f_{rate}$ is the rate of change of ramp frequency and $\dot{f}_{osc}$ is oscillator ramp rate given in the ATDF file.

\subsection{Software architecture and functionalities}
\label{functions}

\begin{figure*}
\centering
\noindent
\includegraphics[width=30pc]{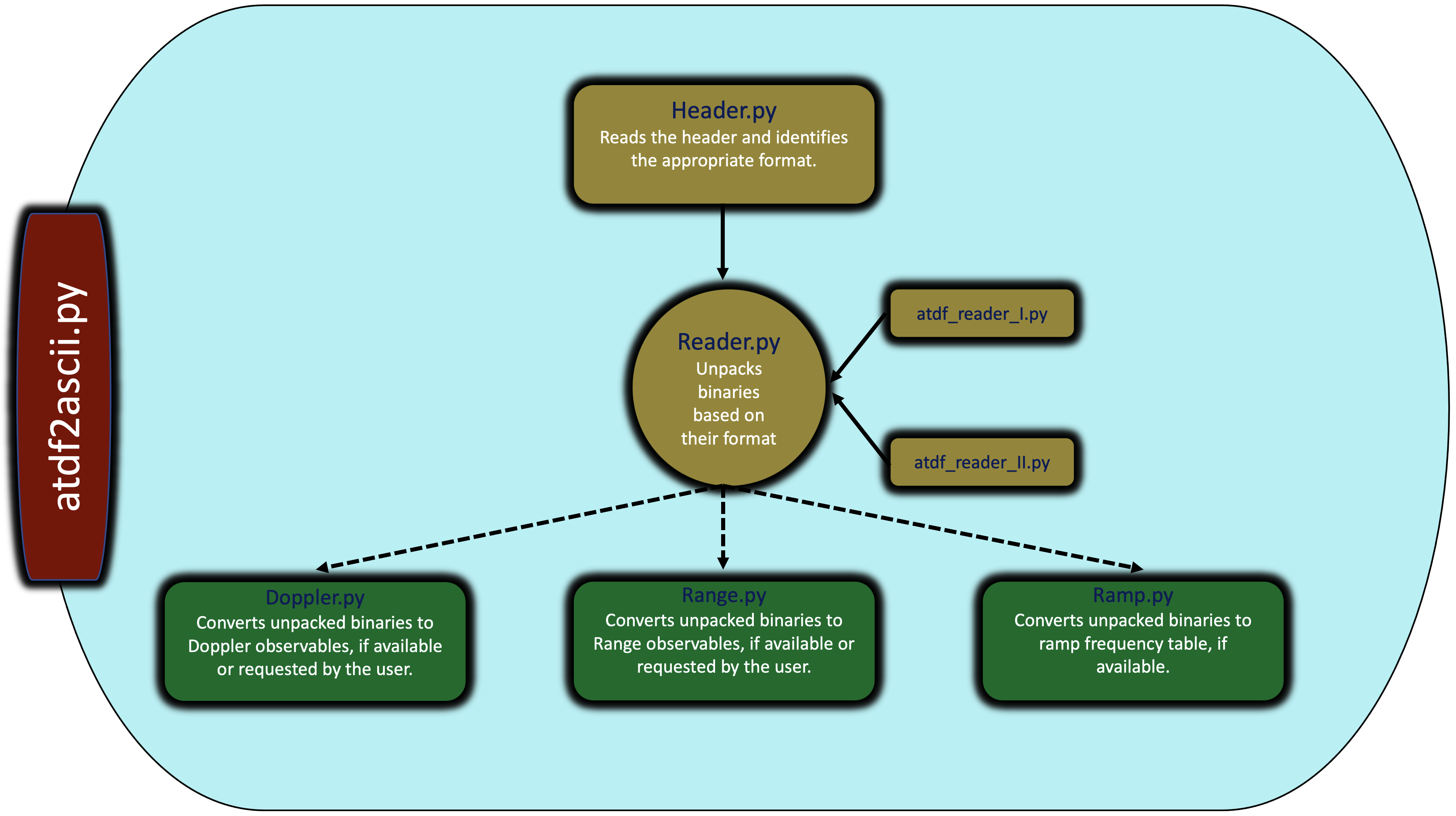}
\caption{Structure of the ATDF software.}
\label{layout}  
\end{figure*}

Figure \ref{layout} lays out the architecture of the ATDF software, where {\fontfamily{qcr}\selectfont atdf2ascii.py} is the main python script to execute the software. Based on the header information of the input ATDF file, software detects the relevant format of the binaries and unpack them using either {\fontfamily{qcr}\selectfont atdf$\_$reader$\_$I.py} or {\fontfamily{qcr}\selectfont atdf$\_$reader$\_$II.py} script. The {\fontfamily{qcr}\selectfont Doppler, Range,} and {\fontfamily{qcr}\selectfont Ramp} classes use these unpacked records and convert them into observables. The software creates two output files upon successful execution: one that contains the Doppler and range observables \added{{(Table \ref{outputT})}}, and the other one contains the ramp frequency history \added{{(Table \ref{rampT})}}. The following command-line options can be used to preprocess an ATDF file: 

\linespread{1}
\begin{verbatim}
       Python atdf2ascii.py -i input_file.tdf [options ...]
\end{verbatim}

where, {\fontfamily{qcr}\selectfont input$\_$file.tdf} is the TRK-2-25 formatted input data file and the options are described as follows: 

\begin{itemize}
   \item {\bf{\fontfamily{qcr}\selectfont --count$\_$time [-c]}}: this option can be used to specify the Doppler count time, $T_c$, in seconds, at which Doppler measurements must be compressed. It accepts one or more values (separated by commas). In the event that more than one count time is specified, the last value in the list will be preferred if the requested count time is not found in the original records. By default, the count time at which the original data is recorded will be used. \added{{ To select the count time, the type of radio science experiment must be considered, and the selection should be made in a way that maintains the key features of the radio signal over the entire count interval. For instance, in experiments analyzing atmospheric dynamics, it may be necessary to use a shorter count time (1 second or less) to account for the low frequency fluctuations in the observed signal. On the other hand, for overly sensitive experiments, such as testing fundamental physics, better sensitivity can be achieved with a longer count time (e.g., 1000 seconds). For further reading, see \cite{Asmar05, Asmar22}}}. 
   
   \item {\bf{\fontfamily{qcr}\selectfont --proc$\_$count [-p]}}: this option limits the number of CPUs used in multi-processing. By default, half of the available CPUs will be utilized.
   
   \item {\bf{\fontfamily{qcr}\selectfont {-}{-}xd1}}: this option precludes the processing of one-way Doppler measurements.
   \item {\bf{\fontfamily{qcr}\selectfont {-}{-}xd2}}: this option precludes the processing of two-way Doppler measurements.
   \item {\bf{\fontfamily{qcr}\selectfont {-}{-}xd3}}: \added{ this option precludes the processing of three-way Doppler measurements.}
   \item {\bf{\fontfamily{qcr}\selectfont {-}{-}xr1}}: this option precludes the processing of one-way Rang measurements.
   \item {\bf{\fontfamily{qcr}\selectfont {-}{-}xr2}}: this option precludes the processing of two-way Rang measurements.
   
\end{itemize}

\begin{table}[H]
\caption{{\added{A column wise description of the output file containing raw observables with ancillary information. Each column of the output file is delimitated by the comma. }}}
\label{outputT} 
\begin{threeparttable}
\begin{tabular}{>{\columncolor[gray]{0.95}}l >{\columncolor[gray]{0.95}}p{4.0cm}  >{\columncolor[gray]{0.95}}p{8.0cm}l}
\hline
& & \\
Column & Output type & Description \\ 
& & \\ \hline\hline
1 & Time tag & This is the measurement time tag. For Doppler, it refers to the midpoint of the count interval at the receiving station in UTC. \\
2 & Data type & The type of radio science measurement. It could be either, 1-Way-Range, 2-Way-Range, 1-Way-Doppler, 2-Way-Doppler, and 3-Way-Doppler. \\ 
3 & Spacecraft ID & This is the SPICE ID of the spacecraft. \\
4 & Transmitter & The name of the transmitting DSN station. \\
5 & Receiver & The name of the receiving DSN station. \\
6 & Channel & The Doppler channel number. \\
7 & Uplink band & The uplink frequency band (transmitter to the spacecraft). \\
8 & Downlink band & The downlink frequency band (spacecraft to the receiver). \\ 
9 & Exciter band & The exciter band of the receiver. \\
10 & Count time & The Doppler count time in seconds. This could be user's input (see Section \ref{functions}). \\ 
11 & Range-low-component & The lowest component number of the frequency ranging code. \\
12 & Observed & The observed value of the measurement. The value is in Hz for Doppler or Range Units for Range measurements. \\
13 & Reference frequency & The reference frequency of the receiver. \\
14 & Transmitter delay & The time delay at the transmitting station, in nanoseconds.\\
15 & Receiver delay & The time delay at the receiving station, in nanoseconds. \\
16 & Spacecraft delay & The time it takes a spacecraft to receive and transmit a signal. \\ \hline

\end{tabular}%
\end{threeparttable}  
\end{table}

\begin{table}[H]
\caption{{\added{A column wise description of the output file containing ramp frequency history. Each column of the output file is delimitated by the comma. }}}
\label{rampT} 
\begin{threeparttable}
\begin{tabular}{>{\columncolor[gray]{0.95}}l >{\columncolor[gray]{0.95}}p{4.0cm}  >{\columncolor[gray]{0.95}}p{8.0cm}l}
\hline
& & \\
Column & Output type & Description \\ 
& & \\ \hline\hline
1 & Start time & This is start time of the ramp frequency. \\
2 & End time & This is end time of the ramp frequency. \\
3 & Station & The name of the transmitting DSN station. \\
4 & Band & The transmitting frequency band. This could be S, X, or Ka. \\
5 & Frequency & The start frequency of the ramp, in Hz. \\
6 & Frequency rate & The rate of change of the frequency, in Hz/sec. \\
\hline

\end{tabular}%
\end{threeparttable}  
\end{table}

\subsection {Mars Global Surveyor ATDF data}
\label{mgs}

\section{Results}
\label{}

Here we are processing two independent ATDF data files: one is from NASA Magellan mission when the spacecraft was in orbit around Venus and the other is from NASA Mars Global Surveyor (MGS) mission when the spacecraft was in orbit around Mars.

\subsection {Magellan ATDF data}
\label{mgn}

The Magellan mission was launched by NASA in 1989 to map the surface of Venus and estimate its gravitational field \added{{\citep{Saunders1990, Saunders1991, Konopliv1991}}}. The Magellan spacecraft reached Venus in August 1989 and began collecting data from September 1990 (cycle 1) through October 1994 (cycle 5). The ATDF data are the primary closed-loop radio science data acquired by the spacecraft and are available through the Geoscience Node of the Planetary Data System. 
For illustration purposes, we randomly selected the ATDF file\footnote{\url{https://pds-geosciences.wustl.edu/mgn/mgn-v-rss-1-tracking-v1/mg\_2601/TDF/}} and preprocessed it through our software.

\begin{figure*}
\centering
\noindent
\includegraphics[width=30pc]{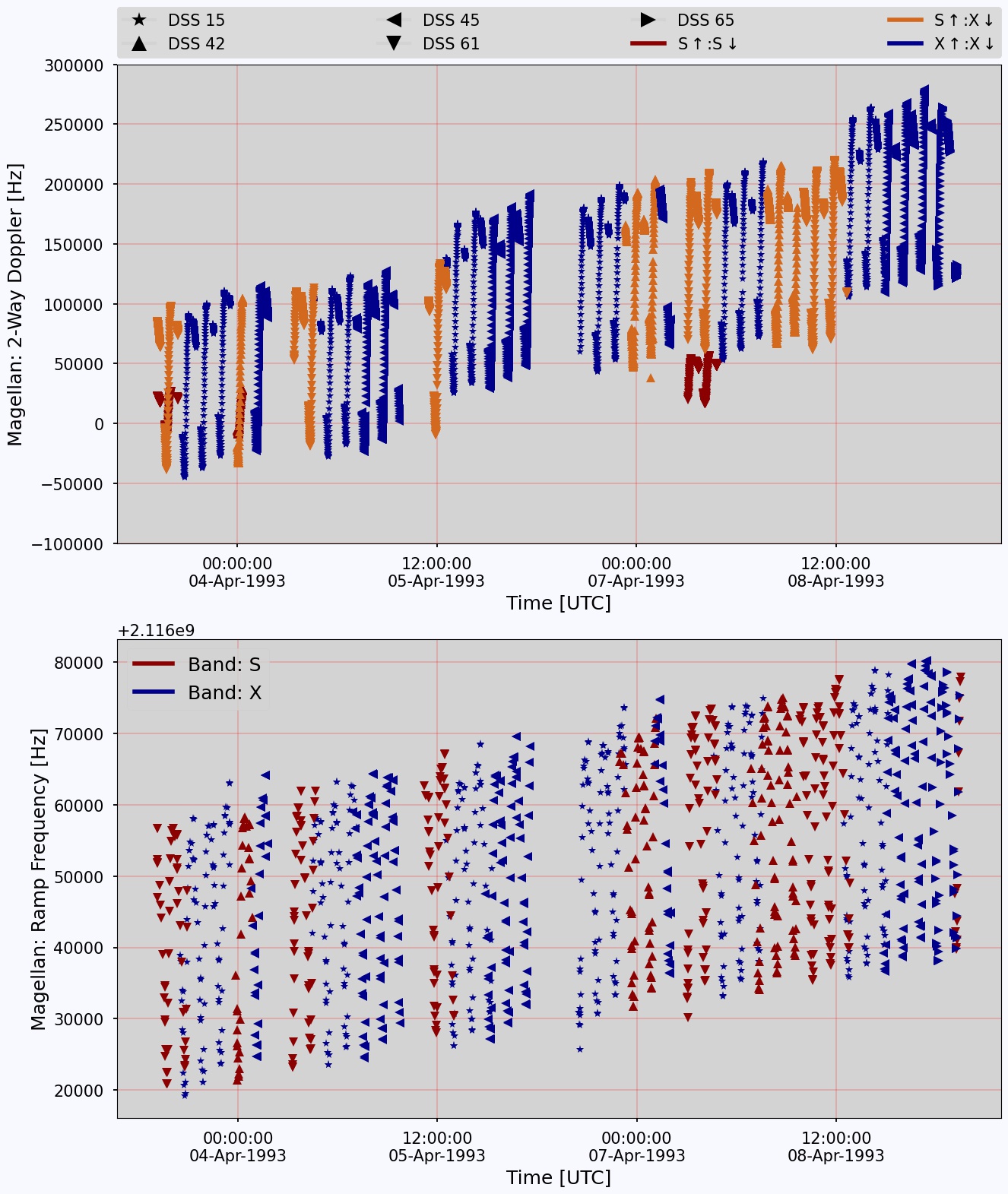}
\caption{A two-way Doppler observable (upper) and the history of the ramp frequency (lower) derived from raw ATDF data are shown in this figure.  \deleted{The left part of the} Figure shows Magellan observables when the spacecraft was orbiting Venus. \deleted{while the right part shows Mars Global Surveyor observables when the spacecraft was orbiting Mars.}}
\label{mgnData}  
\end{figure*}

The Magellan data file consisted of records that spanned from one-tenth to 60 seconds. In \deleted{the left panel of} Figure \ref{mgnData}, we show the raw values of the Doppler observables and receiver ramp frequencies that we derived from the ATDF file. These Doppler observables were constructed by using count times, $T_c$, of 10 and 60 seconds with the following command:

\linespread{1}
\begin{verbatim}
       Python atdf2ascii.py -i 3093099a.tdf -c 10, 60
\end{verbatim}

The colors in the figure indicate different sets of uplink and downlink bands, while the symbols represent the DSN stations. Two files were produced from the above command: one contains the observables with accompanying information such as time tags, DSN stations, uplink and downlink bands, and delays \added{{(Table \ref{outputT})}}, and the other contains the time history of the ramp frequencies \added{{(Table \ref{rampT})}}. These measurements are vital to gravity investigations, allowing the reconstruction of the spacecraft's orbit and estimating the geophysical attributes of the body.

\begin{figure*}
\centering
\noindent
\includegraphics[width=30pc]{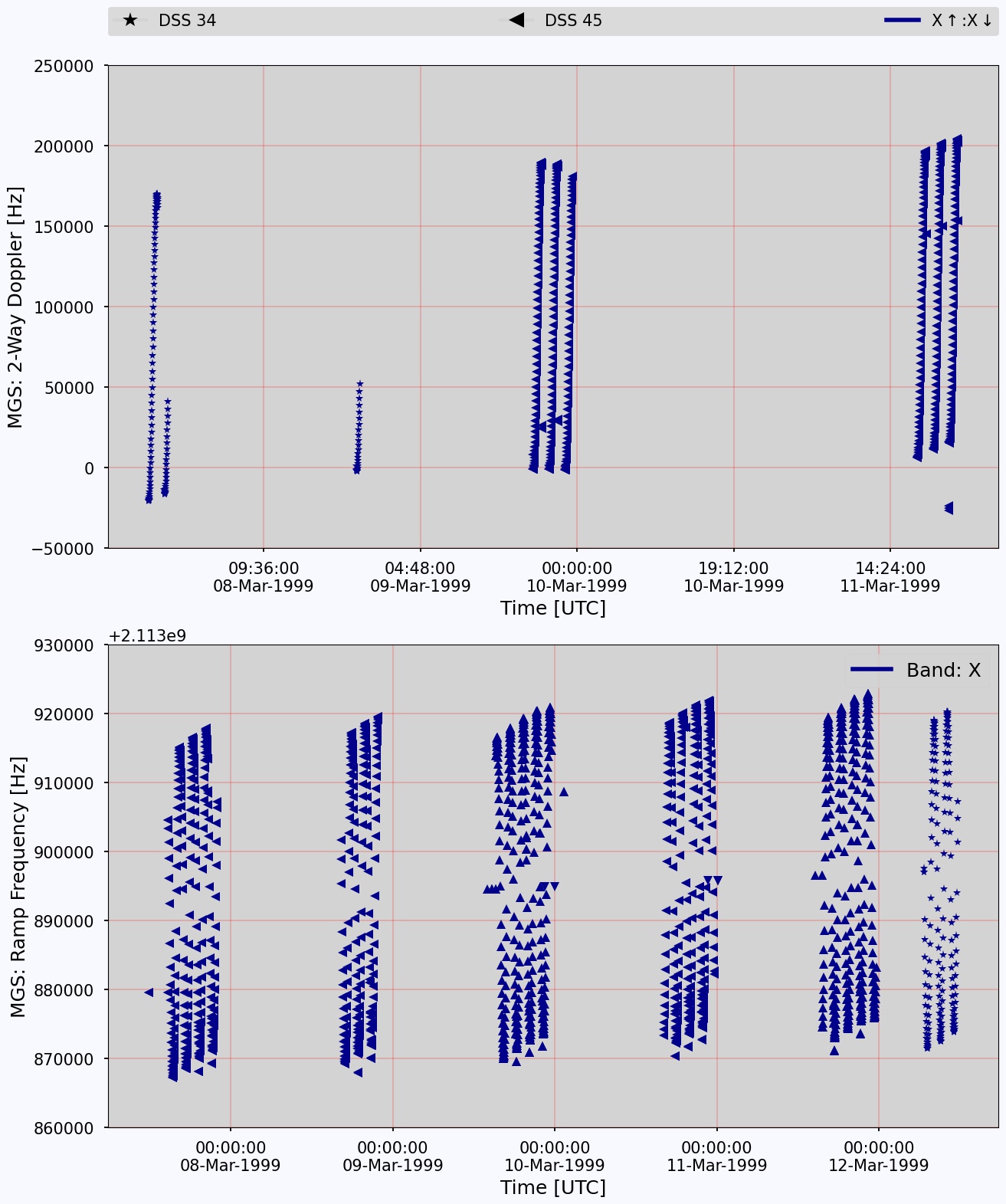}
\caption{A two-way Doppler observable (upper) and the history of the ramp frequency (lower) derived from raw ATDF data are shown in this figure.  \deleted{The left part of the figure shows Magellan observables when the spacecraft was orbiting Venus, while the right part} \added{{ Figure}} shows Mars Global Surveyor observables when the spacecraft was orbiting Mars.}
\label{mgsData}  
\end{figure*}

\begin{table}[H]
\caption{{\added{An overview of the TRK-2-25 closed-loop radio science data that were used in the examples. 
A links to the raw data files along with detailed PDS3 metadata are provided in the footnotes. 
Ascii outputs obtained after preprocessing of these data are given in the Example directory.}}}
\label{tab:metadata} 
\begin{threeparttable}
\begin{tabular}{>{\columncolor[gray]{0.95}}l >{\columncolor[gray]{0.95}}p{5.0cm}  >{\columncolor[gray]{0.95}}p{5.0cm}l}

\hline
Mission &
  Magellan &
  Mars Global Surveyor \\ \hline
Data Files & Raw\tnote{1}, Metadata\tnote{2} & Raw\tnote{3}, Metadata\tnote{4} \\
Start Time &
  1993-04-03T09:09:50 UTC &
  1993-04-03T09:09:50 UTC \\ 
Stop Time &
  1999-03-07T11:46:54 UTC &
  1999-03-12T11:45:00 UTC \\ 
Data Type &
  one-way Doppler, two-way Doppler, three-way Doppler, ramp &
  one-way Doppler, two-way Doppler, three-way Doppler, one-way range, two-way range, ramp\\
Transponder Freq. &
  2297963786.0 Hz &
  2297222222.0 Hz \\
DSN Stations &
  15, 42, 45, 61, 65 &
  15, 34, 45, 54 \\
Bands &
 S, X &
  S, X \\
  \hline
\end{tabular}%
  \begin{tablenotes}
    \item[1] \url{https://pds-geosciences.wustl.edu/mgn/mgn-v-rss-1-tracking-v1/mg_2601/TDF/3093099a.tdf}
    \item[2] \url{https://pds-geosciences.wustl.edu/mgn/mgn-v-rss-1-tracking-v1/mg_2601/TDF/3093099a.lbl}
    \item[3] \url{https://pds-geosciences.wustl.edu/mgs/mgs-m-rss-1-map-v1/mors_0402/tdf/9066071a.tdf}
    \item[4] \url{https://pds-geosciences.wustl.edu/mgs/mgs-m-rss-1-map-v1/mors_0402/tdf/9066071a.lbl}
  \end{tablenotes}
\end{threeparttable}  
\end{table}

\subsection {Mars Global Surveyor ATDF data}
\label{mgs}

Mars Global Surveyor (MGS) was a NASA mission to study Mars, launched in 1996. The primary phase of the mission began in 1999, and a wealth of data were sent back to Earth until 2006 \added{{\citep{Albee2001,Genova2016}}}. Similar to Magellan, ATDF was the primary closed-loop product of the MGS spacecraft. At the PDS, however, data files with ATDF format were not often archived; instead, a simpler form of closed-looped data, Orbit Data Files (ODFs)\cite{odf}, were archived\footnote{\url{https://pds-geosciences.wustl.edu/mgs/mgs-m-rss-1-map-v1/}}. These ODFs were created from ATDFs and provide direct access to the observables (similar to the output from the proposed software).

Depending on the investigation, MGS ATDF data was recorded down to one-tenth of a second. \deleted{The right panel of} Figure \ref{mgsData}, shows the raw values of the Doppler observables and receiver ramp frequencies. Using a fixed count time, $T_c$, of 60 seconds, the following commands were used to construct the Doppler observables:

\linespread{1}
\begin{verbatim}
       Python atdf2ascii.py -i 9066071a.tdf -c 60 --xd1
\end{verbatim}
 
where {\fontfamily{qcr}\selectfont {-}{-}xd1} flag excluded one-way Doppler observables.  

\subsection {Validation}
\label{valid}

To validate the constructed values of the raw observables, we compared them to predicted values. Following the \replaced{{realistic} }{relativistic} Moyer's formulation, the predicted values were obtained using positions and velocities of the spacecraft and bodies of the solar system. We assessed these positions and velocities through archived ephemerides of the spacecraft\footnote{\url{https://naif.jpl.nasa.gov/naif/data\_operational.html}} and DE438 planetary ephemerides\cite{FolknerDE438}. 

\begin{figure}
\centering
\noindent
\includegraphics[width=30pc]{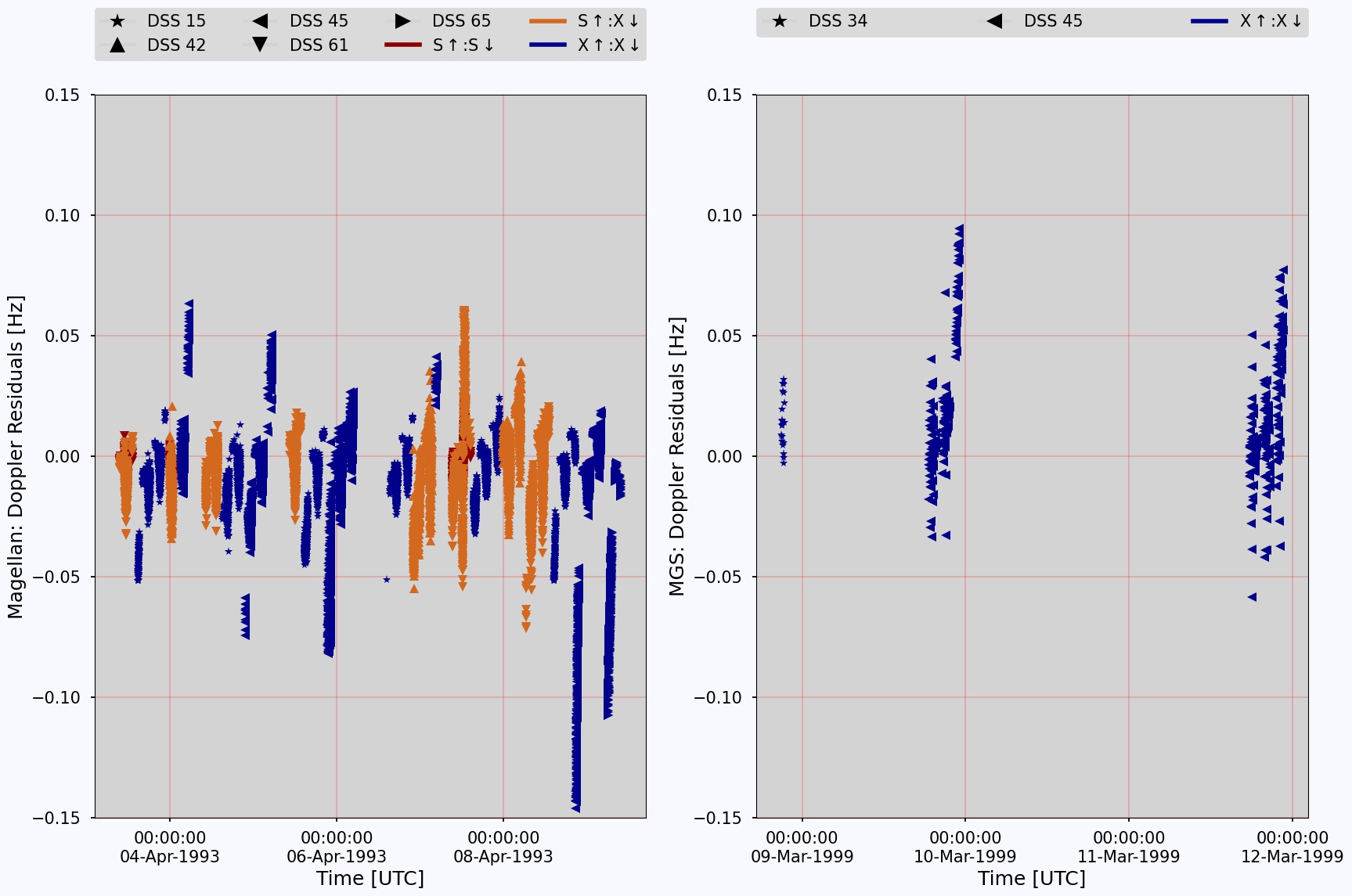}
\caption{The Doppler residuals of Mars Global Surveyor (left) and Magellan (right) spacecraft. These residuals were computed by comparing the raw and predicted observables (see Section \ref{valid}).}
\label{resd}  
\end{figure}

Figure \ref{resd} shows the Doppler residuals for the two spacecraft. The root-mean-square values of residuals for Magellan and MGS spacecraft are 22 mHz and 26 mHz, respectively. The remaining trends in the residuals are presumably due to errors in the spacecraft trajectory and a mis-modeling of the measurement model, which does not account for spacecraft attitudes, media delays, antenna positions, etc. 
Nevertheless, any error in the constructed values of the raw observables would have been straightforward to distinguish, as those errors would usually be associated with offsets. Therefore, the residual values shown in Figure \ref{resd} are indicative of an accurate retrieval of the raw observables.

\section{Impact}
\label{}

Radio science data are the primary data sets used for deep space navigation. Over the past half-century, these data have been used for deep space exploration to reveal the geophysical properties of solar system bodies. The ATDF format was deprecated by the DSN in the early 2000s and replaced with the TNF \added{{TRK-2-34 format\cite{tnf}}} to accommodate the evolution of the radio science subsystem. It was until then the primary closed-loop radio science data were used widely for navigation, science, and creating a simpler closed-loop data format, ODF\cite{odf}.  

These historical data are still as valuable today as they were when they were first acquired. Given the current advancement in computing power, a better understanding of the solar system, and very precise software provide a valid reason to re-analyze such data sets. NASA, for instance, has announced two Discovery missions to visit Venus by late 2020. As such, re-analyzing Magellan radio science data with modern tools would be vital to the planning and design of these missions.

Despite being valuable science data, there are no open-source tools for preprocessing complex radio science data. The tool presented in this paper enables users to preprocess closed-loop data and create observables that can be further analyzed with modern orbit determination software.  We believe that this tool will make      historical radio science data more accessible to users outside of radio science community. We also expect that it will become a starting point for those who are interested in analyzing such type of data in the future.

\section{Conclusions}

This paper describes a Python-based tool that can preprocess the closed-loop radio science data packed in the ATDF format. In addition to producing Doppler and range observables, this tool also produces information such as receiver ramp frequency history, uplink and downlink bands, DSN stations, delays, etc., and writes to an ASCII table. Using the command-line option {\fontfamily{qcr}\selectfont -c}, the user may produce Doppler observables at either a fixed or variable count time. A comparison was made between derived Doppler observables and the predicted observables for validation. This comparison shows no apparent offset in the residuals, verifying the success of creating Doppler observables. 

The observables, along with ancillary information, are essential for reconstructing spacecraft orbits and conducting radio science experiments. To our knowledge, there are no open-source tools available for preprocessing closed-looped radio science. Therefore, this tool increases the usability of historical and valuable data and allows users to process them using modern techniques and software.

\section*{Declaration of Competing Interest}

The authors declare that they have no known competing financial interests or personal relationships that could have appeared to influence the work reported in this paper.

\section*{Acknowledgements}
\label{}
\added{{ The author wishes to thank all three reviewers for their insightful remarks and suggestions concerning this manuscript and software.}}
This research was funded by NASA Cooperative Research Notice Award 80NSSC22M0023.

\bibliographystyle{elsarticle-num} 
\bibliography{reference}

\end{document}